\documentclass[10pt]{iopart}
\pdfoutput=1
\usepackage{bmpsize}
\usepackage{graphicx,color}								
\usepackage{amssymb}
\expandafter\let\csname equation*\endcsname\relax
\expandafter\let\csname endequation*\endcsname\relax
\usepackage{amsmath}
\usepackage{array}
\usepackage{subfigure}
\usepackage{upgreek}
\usepackage[usenames,dvipsnames,svgnames,table]{xcolor}
\newcommand{\jh}[1]{{\color{black} #1}}
\newcommand{\paddyspeaks}[1]{{\color{black} #1}}

\begin{document}

\title{The Devil is in the details: Pentagonal bipyramids and dynamic arrest}

\author{James E. Hallett$^{1,2}$, Francesco Turci$^{1}$ and C. Patrick Royall$^{1,2,3}$*}

\address{$^1$H.H. Wills Physics Laboratory, Tyndall Avenue, Bristol, BS8 1TL, UK}
\address{$^2$Centre for Nanoscience and Quantum Information, Tyndall Avenue, Bristol, BS8 1FD, UK}
\address{$^3$School of Chemistry, University of Bristol, Cantock's Close, Bristol, BS8 1TS, UK}
\ead{*paddy.royall@bristol.ac.uk}

\begin{abstract}

Colloidal \jh{suspensions} have long been studied as a model for atomic and molecular systems, due to the ability to fluorescently label and individually track each particle, yielding particle-resolved structural information. This allows various \jh{local} order parameters to be probed that are otherwise inaccessible for a comparable molecular system. For phase transitions such as crystallisation, appropriate order parameters which emphasise 6-fold symmetry are a natural choice, but for vitrification the choice of order parameter is less clear cut. Previous work has highlighted the importance of icosahedral local structure as the glass transition is approached. However, counting icosahedra or related motifs is not a continuous order parameter in the same way as, for example, the bond-orientational order parameters $Q_{6}$ and $W_6$. In this work we investigate the suitability of using pentagonal bipyramid membership, a structure which can be assembled into larger, five-fold symmetric structures, as a finer order parameter to investigate the glass transition. We explore various structural and dynamic properties and show that this new approach produces many of the same findings as simple icosahedral membership, but we also find that large instantaneous displacements are often correlated with significant changes in pentagonal bipyramid membership, and unlike the population of defective icosahedra, the pentagonal bypyramid membership and spindle number do not saturate for any measured volume fraction, but continues to increase.

\end{abstract}

%\pacs{61.20.-p; 64.70.Dv; 61.20.Gy}

\maketitle

\section{Introduction}

The dramatic slowdown in dynamics in supercooled liquids approaching the glass transition remains unexplained \cite{berthier2011,cavagna2009}. By contrast, in the case of freezing, solidification may be linked to the local crystalline structure assumed by the constituent particles. Central to the conundrum of the glass transition then, is how the viscosity can increase by \emph{14 orders of magnitude} with little apparent change in structure \cite{royall2015physrep}. A wide variety of theoretical approaches have been advanced to address this challenge, for which the reader is referred to a number of excellent reviews  
\cite{berthier2011,cavagna2009,debenedetti,dyre2006,stillinger2013} and shorter perspectives \cite{berthier2016pt,biroli2013,debenedetti2001,ediger2017,ediger2012,berthier2019}.

Theories with particular relevance from the point of view of local structure include Adam-Gibbs \cite{adam1965} and Random-First-Order-Transition (RFOT) theories \cite{lubchenko2007}, and geometric frustration \cite{tarjus2005}. Adam-Gibbs and RFOT posit a drop in the \emph{configurational} entropy of the system, which should become sub-extensive at some finite temperature, \jh{corresponding} to the Kauzmann transition \cite{kauzmann1948}, i.e. a so-called ideal glass. Recently, considerable progress has been made in approaching this putative state \cite{royall2018jpcm}, and drops in configurational entropy determined from order-agnostic measures of structure have been obtained in experiments on colloids \cite{gokhale2016,gokhale2016jsm,williams2018,hallett2018} computer simulation of structural glasses \cite{berthier2017pnas,berthier2019}, 
\jh{and even some spin glasses \cite{baityjesi2019}.}
On the other hand, building on the idea that five-fold symmetric \emph{locally favoured structures} are ``abhorrent to crystallisation'', as they cannot tile 3D space \cite{frank1952}, geometric frustration relates the slowdown in dynamics in Euclidean space to a presumed transition to a state of icosahedra in some non-frustrated space \cite{tarjus2005}.

In testing out some of these ideas, particular attention has focussed on hexagonal order in 2D, where the situation is clearer (there appears to be mainly one form of order) and visualisation is straightforward. Growing domains of hexagonal order have been found in colloid experiments \cite{yunker2009,tamborini2015,gokhale2016}, granular experiments \cite{candelier2010,watanabe2011} and computer simulation \cite{kawasaki2007}, and related to a 2D interpretation of geometric frustration \cite{sausset2010}. While some controversy exists over the connection between order-agnostic \emph{point-to-set} lengths and explicit determination of hexagonal symmetry \cite{russo2015,yaida2016}, this now appears to have been resolved by careful consideration of the fluctuation of thermodynamic quantities in the cavities (small local regions) probed in calculations of the point-to-set length \cite{tah2018}. However, significant differences have been found in the nature of the dynamics between two and three dimensions in simulations \cite{flenner2015}, and colloid experiments \cite{vivek2017}, moreover evidence has recently emerged that in 2D, there may be no finite-temperature glass transition \cite{berthier20182d}.

One might therefore reasonably enquire what colloid experiments (and computer simulations) might tell us of the glass transition in the three-dimensional world in which we live. \paddyspeaks{While correlation between local structures and dynamical properties has been found \cite{royall2015physrep,wu2013,marin2018,dasgupta2019,marin2019}, a particularly interesting feature would be
}%Key to this issue is 
the co-incidence of  dynamic and structural lengthscales \paddyspeaks{at sufficient supercooling} \cite{royall2015physrep,karmakar2014}. 
Such a coming together of lengthscales would provide significant support for the RFOT/Adam-Gibbs picture \cite{berthier2011,karmakar2014}. However, until recently, the vast majority of work has found that dynamic lengthscales, (defined often through the so-called $\xi_4$ measure, which presumes an Ornstein-Zernike fit to the 4-point structure factor $S_4(q,t)$ \cite{lacevic2003}) do \emph{not} scale with a variety of structural lengths \cite{royall2015physrep}. Exceptions include 2D work \cite{kawasaki2007,tanaka2010} (but note the caveats above), alternative definitions of the dynamic lengthscale \cite{harrowell2011,karmakar2014,dunleavy2015,royall2017,pastore2017} and experimental work on colloids \cite{leocmach2012}. A key observation is that \jh{the previous work has} largely been limited to the first four decades of supercooling, and thus it is hard to conclude anything directly about the glass transition itself.

%\begin{figure}
%\centering
%\includegraphics[width=0.9\textwidth]{figXiThis}
%\caption{\textbf{Lengthscales approaching the glass transition.}
%Structure-based theories such as Adam-Gibbs and RFOT (and geometric frustration) imagine dynamic and structural lengthscales that grow concurrently (a). However, the majority of studies access only the weakly supercooled regime, to around the mode-coupling transition, $T_\mathrm{MCT}$. In the majority of cases, the dynamic lengthscale increases faster than the static (structural lengthscale) (b), suggesting that some change in the scaling would be required at deeper supercooling \cite{royall2015physrep}. 
%}
%\label{figXiThis}
%\end{figure}

Another approach is to directly interrogate a specific theory, and recently, the theory which makes the most explicit reference to local structure, geometric frustration, has been investigated \cite{turci2017}. In curved space, the expected sharp crossover to a state of icosahedra is found, and for small amounts of frustration, indeed the structural and dynamic lengthscales couple. However, the increase in frustration upon uncurving the space towards the Euclidean case, leads to a decoupling in dynamic lengthscales and lengthscales related to icosahedra. Therefore in the dynamical regime accessible to computer simulation and colloid experiment, where higher-order structure can be rigourously defined and compared with dynamic lengthscales, we conclude that growth of five-fold symmetric domains does not, by itself, explain the slowdown in dynamics. Moreover, it has been shown that local structure is %very 
sensitive to the system in question \cite{hocky2014,jack2014,royall2015,coslovich2018}, although it should be possible to define a locally favoured structure or structure\emph{s} for a given system \cite{royall2015}, using energy minimisation \cite{doye1995} or morphometric methods in the case of systems without attractions \cite{robinson2018}.

Given the limitations of geometric frustration -- tested in the %``ideal'' 
Wahnstr\"{o}m binary Lennard-Jones model, known for its high prevalence of five-fold symmetry \cite{coslovich2007,malins2013jcp,royall2015} -- one might reasonably enquire as to \jh{its} use as a metric in systems undergoing a glass transition. On the other hand, local structure \emph{has} been shown to drive a structural-dynamical phase transition in trajectory space \cite{speck2012,turci2017,turci2018epje}, which has found experimental verification \cite{pinchaipat2017}. In the Wahnstr\"{o}m model and hard spheres, \cite{turci2018epje,pinchaipat2017}, this is driven via a bias on five-fold symmetric structures.  Further, direct evidence in support of the role of five-fold symmetry has been obtained using smaller colloids to access an unprecedented dynamic range in particle-resolved studies \cite{hallett2018}.

While real-space measurements such as particle-resolved confocal microscopy and computer simulations can be used to identify local structure in dense colloidal suspensions at the microscale \cite{royall2015physrep,hunter2012,yunker2014,zhang2016}, tantalising insights of this mechanism have also been obtained using scattering methods \cite{liu2016,liu2017,wette2009,wochner2009,dicicco2003,hirata2010,hirata2013}. Nevertheless, despite the presence of five-fold symmetric structures in supercooled systems and their correlation with slow dynamics, there is no universally accepted order parameter to describe the onset of this five-fold order.

For some phenomena, order parameters naturally emerge: for crystallisation order parameters that prioritise 6-fold symmetry are obvious \cite{tanaka2010}, while in active matter systems, it can be informative to consider polarisation of mobility due to cooperative motion of the constituent particles \cite{bricard2013}. For an amorphous system, the solution is less clear. Various approaches have been used to address this issue, such as considering structural overlaps \cite{berthier2013} but one attractive option is to construct an order parameter based on the detection of local structure. We recently demonstrated that local structure plays a key role in the behaviour of colloidal liquids at unprecedented degrees of dynamic slowdown \cite{hallett2018}, noting the significance of defective icosahedra \jh{(regular icosahedra missing three neighbouring particles from a face)} \cite{pinchaipat2017} and regular icosahedra in the development of slow dynamics. In this work we evaluate a new order parameter: the occupancy of pentagonal bipyramids. This is an appealing alternative because both defective icosahedra and regular icosahedra can be constructed from pentagonal bipyramids. Therefore by identifying the local occupancy of these pentagonal bipyramids, one can identify finer changes in the structure as the metastable liquid becomes more and more dominated by local icosahedral order.%one can identify structure that varies \textit{continuously} from liquid-like to highly icosahedral. 

In this work dense suspensions of colloidal hard spheres were prepared in rectangular cells for observation by stimulated emission depletion microscopy. Particle coordinates were tracked and their structure and dynamics were subsequently interrogated by considering the role of the pentagonal bipyramid structure. By studying this new data and re-evaluating data from our previous work that spans nearly seven decades of dynamic slowdown \cite{hallett2018}, we show that the relationship between pentagonal bipyramid occupancy and slow dynamics is comparable to that of icosahedra and defective icosahedra. However, we also show that large instantaneous displacements often coincide with large jumps (either positive or negative) in pentagonal bipyramid occupancy -- while defective icosahedra occupancy remains the same. We also show that the average spindle number and pentagonal bipyramid occupancy number continue to increase with supercooling unlike simple membership of defective icosahedra, which saturates at densities comparable to the mode-coupling crossover due to geometric constraints \cite{hallett2018}.

\section{Methods}

\subsection{Experimental procedure}
Fluorescent, core-shell index and density matched poly(methyl methacrylate) (PMMA) particle suspensions were prepared for microscopy observation \cite{hallett2018}. Briefly, PMMA colloidal particles ($\sigma$ = 540 nm, 8\% polydispersity) were synthesised with a rhodamine-dyed fluorescent core and non-fluorescent shell and transferred into a mixture of cis-decalin and cyclohexylbromide. This mixture matches the refractive index and density of PMMA and, with the addition of tetrabutylammonium borate salt to screen any surface charges, can effectively reproduce hard-sphere behaviour. Stock suspensions were prepared at high volume fraction by centrifugation at elevated temperatures and removal of a known quantity of solvent. Samples were subsequently prepared by filling glass capillaries (approx. 100 micron thickness) with the colloidal suspension and sealing with epoxy glue and were either imaged straight away or allowed to equilibrate for up to several weeks.

3D particle resolved measurements were performed using stimulated emission depletion microscopy (STED) with a Leica SP8 microscope, using a white light laser set to 543 nm for excitation and a 660 nm depletion laser. Imaging volumes were typically (10-15 $\mu$m)$^{3}$ and sampling interval of %rate 
%between 
10 -- 1200 seconds between frames depending on volume fraction. Total sampling times were up to 48 hours. Raw data were subsequently processed using Huygens deconvolution software and particle coordinates were obtained using tracking algorithms \cite{leocmach2013}. 
The structural relaxation time $\tau_{\alpha}$ was obtained from the intermediate scattering function (ISF) for wavevector $q$, which here was taken to correspond to a particle diameter ($q \sim 2 \pi \sigma^{-1}$), close to the main peak in the static structure factor. The long-time tail of the ISF was fitted with a stretched exponential with time constant $\tau_{\alpha}$ and stretching exponent $b$. Structural relaxation times as a function of volume fraction were then fitted using the Vogel-Fulcher-Tammann (VFT) relationship:

\begin{equation}
\tau_{\alpha}\left( \phi \right)=\tau_{\infty}\mathrm{exp}\left[ \frac{A}{\left(\phi_{0} - \phi \right) ^{\delta}}\right]
\label{eqVFTZ}
\end{equation}
\noindent 
where $\tau_{\infty} $ is the relaxation time in a dilute system, $\phi_{0}$ is the the volume fraction at which the relaxation time would diverge, $A$ is a measure of the fragility and $\delta$ is an exponent typically set to one to recover the conventional VFT form.

Where appropriate, figures are plotted as a function of the reduced pressure $Z$ as a control parameter rather than $\phi$, following the arguments of Berthier and Witten \cite{berthier2009pre}. To calculate this we use the Carnahan-Starling equation of state ($Z_\mathrm{cs}=(1+\phi+\phi^2-\phi^3)/(1-\phi)^3$) which gives good agreement with simulations over the range of volume fractions described here \cite{berthier2016swap}.
Local structure was assigned using a modified topological cluster classification \cite{malins2013jcp}, where in addition to assigning different local structures to different particles, the number of specific pentagonal bipyramid clusters or spindles each particle contributed to was also recorded (see below).

\subsection{Structural order parameter}

\begin{figure}
\centering
\includegraphics[width=1\textwidth]{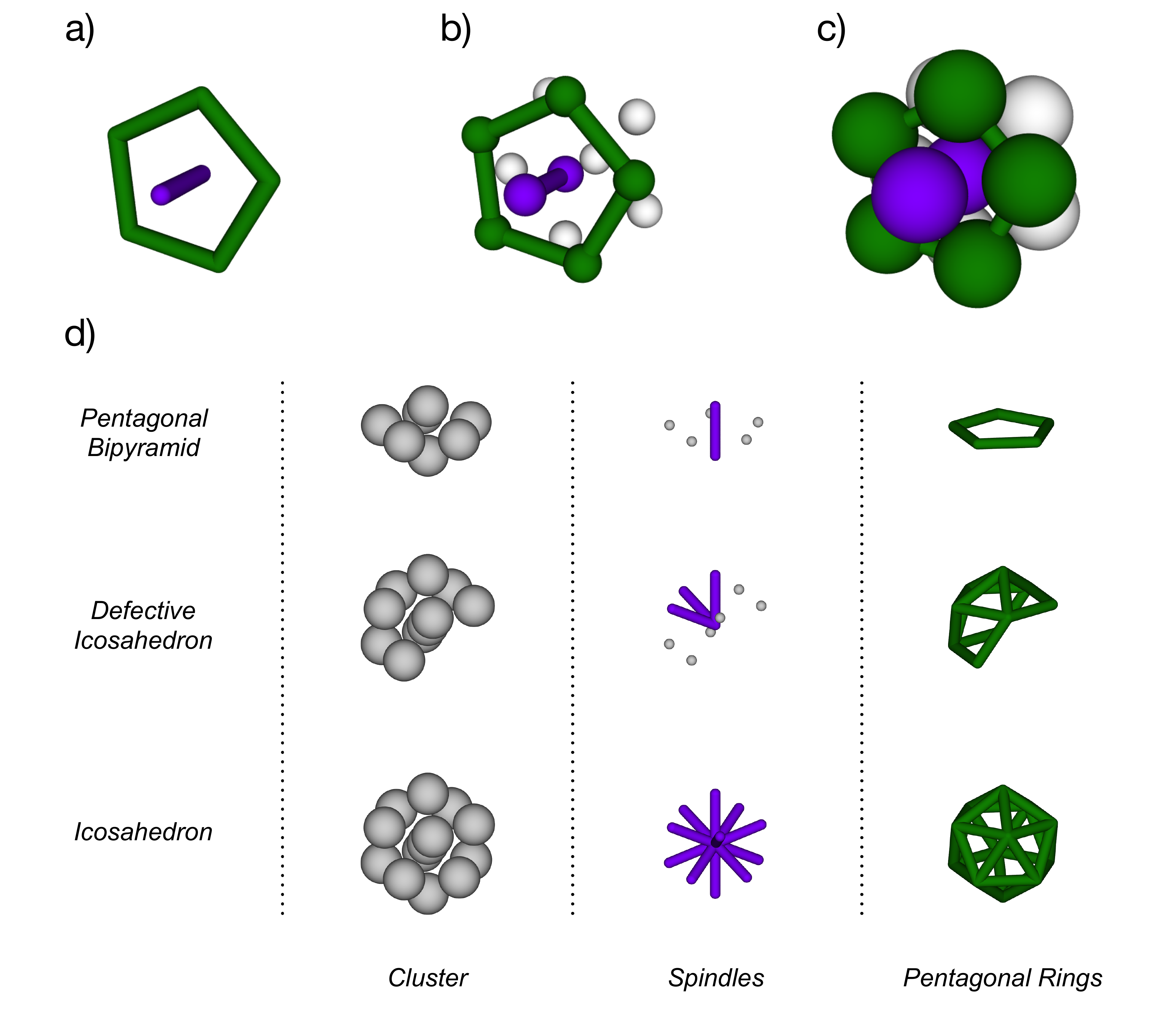}
\caption{\textbf{Pentagonal bipyramids and icosahedra} Rendered pentagonal pyramid bond structure in a) isolation and b) as part of an icosahedron. The spindle is shown in purple and the pentagonal ring is indicated in green. c) Shows a complete icosahedron with one pentagonal bipyramid indicated. It is clear to see that alternative pentagonal bipyramids can be highlighted by simply drawing a spindle from the centre to any particle at the surface, and creating a 5 membered ring from the neighbouring particles. \jh{d) The spindle and pentagonal ring structure for pentagonal bipyramid, defective icosahedron and regular icosehedron clusters. Particles that do not form spindles are also shown to illustrate the position of the spindles in the locally favoured structures.}}
\label{figStructure}
\end{figure}

Previous work has shown that 5-membered rings correlate with \textit{slow} dynamics in a model glassformer \cite{dunleavy2015}. While these rings are necessary to construct icosahedral structures, pentagonal rings alone are not enough. The rings must also be \jh{crossed} by a spindle - generating a pentagonal bipyramid (see figure \ref{figStructure}a). Therefore we choose to investigate the suitability of the pentagonal bipyramid as an order parameter for dynamic arrest. Consider a single icosahedron. The central particle forms part of 12 pentagonal bipyramids, where each spindle is defined by the vector between the central particle and any shell particle, and the five-fold ring is the ring of particles surrounding this spindle (figure \ref{figStructure}b). Therefore the central particle can be a member of 12 pentagonal bipyramids and form 12 spindles. However, if we consider a shell particle, it only forms one spindle (joining the central particle) but can also contribute to 5 rings (for its five shell neighbours), so this particle will form \jh{1 spindle and be a member of 6 pentagonal bipyramids (1 as a spindle former and 5 as a ring member)}. If we now consider an icosahedron in a configuration of particles, it is easy to see how other numbers of pentagonal bipyramids and spindles can be obtained, by considering incomplete icosahedra (so-called defective icosahedra) (figure \ref{figStructure}d) and particles forming part of a network of icosahedra. We thus obtain two order parameters for each particle: The number of pentagonal bipyramids it occupies and the number of spindles it forms, which we denote $O_{\mathrm{PBP}}$ and $O_{\mathrm{S}}$ respectively. This characterisation serves three key purposes: It provides a measure of ``icosahedral quality'', in that it can encapsulate both regular and defective icosahedral order. It also provides a sense of local icosahedral environment, in that large occupancy values are not possible for isolated Locally Favoured Structure (LFS) and instead require a structured domain. Finally, it distinguishes particles by their position in a particular LFS - that is to say a particle in the centre of an icosahedron is distinct from those in the shell, rather than captured within the same characterisation. The relationship between icosahedral structure and pentagonal bipyramids has been explored previously in simulations. By biasing towards the formation of pentagonal bipyramids, crystallisation was suppressed \cite{taffs2016} and icosahedral structures were promoted \cite{taffs2016,carter2018}.
\jh{In order to detect the degree of pentagonal bipyramidal structure, we use a modified topological cluster classification (TCC) scheme. The TCC is a method for identifying local structure in particle configurations. The algorithm detects clusters within a configuration that have previously been identified as “ground state clusters” – clusters that form from a small number (up to 15) of isolated particles that interact via various pair potentials. In the conventional TCC scheme, the algorithm creates a neighbour network for the configuration and identifies three-, four- and five-membered shortest path rings. The rings are further categorised by identification of common neighbours to all ring particles. If there is only one neighbour, this yields a tetrahedron, square pyramid or pentagonal pyramid. If there are two neighbours, by definition (as nearest neighbours to all particles in the ring) these must be above and below the ring, forming  a spindle \cite{malins2013jcp}. This generates a set a of basic clusters: the triangular bipyramid, octahedron and pentagonal bipyramid for the three-, four- and five-membered rings respectively. More complex clusters can form from more particles, but crucially these are still identified through the combination of one or more of the basic clusters and/or additional single particles. The approach adopted here differs at this key step. Instead of using the basic clusters to identify more complex clusters, we instead focus solely on the pentagonal bipyramid cluster. Adjacent clusters can share particles and so individual particles can be used to construct multiple five-membered rings or spindles with different sets of neighbouring particles. We thus record the number of spindles and complete pentagonal bipyramids that each particle contributes toward ($O_{\mathrm{S}}$ and $O_{\mathrm{PBP}}$ respectively). It should be noted that the conventional TCC scheme does use properties such as shared spindles and shared ring-members as part of the logic flow to identify compound clusters, but crucially while a particle has to form a specific number of five-membered rings or spindles to form a particular cluster, it is also possible for it to form \textit{more} five-membered rings and spindles but \textit{not} form any extra compound clusters, and so the TCC does not detect this. For example, assuming no more complex structure is generated, no distinction is made between a particle in a single isolated pentagonal bipyramid, and one shared between two pentagonal bipyramids. It is this distinction that motivates the present study.}

\section{Results}
In this section we will investigate real-space particle resolved data and determine the role of pentagonal bipyramid occupancy in %glassformer 
\jh{the structure and dynamics of supercooled hard spheres.} We first show the spatial distribution of pentagonal bipyramid %PBP 
structural order for a range of state points, revealing spatial heterogeneity which can be described by an increasing lengthscale with volume fraction. We also show that the distribution of PBP and spindle occupancy increases with volume fraction. %density.

We then consider the relationship between PBP membership and slow dynamics in two different ways: We first consider the correlation between mobility and persistence of membership in PBP during trajectories \jh{and} %. We 
then discuss the relationship between instantaneous displacements and changes in PBP membership.

%renders of configurations
%lengthscales

\subsection{Global structure}

\begin{figure}
\centering
\includegraphics[width=0.99\textwidth]{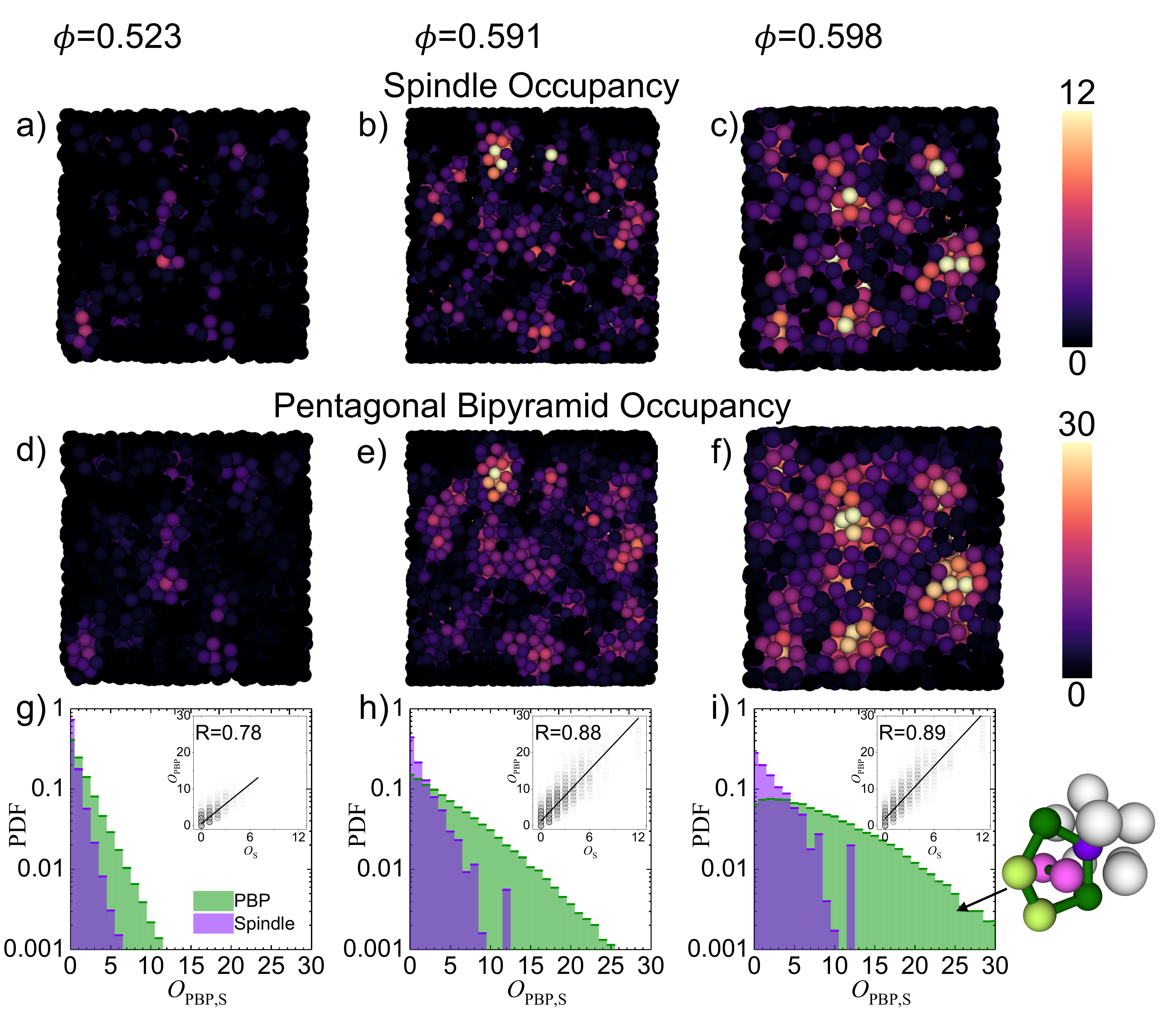}
\caption{\textbf{Five-fold symmetry in configurations}
Rendered configurations of local structure and their distributions at different volume fractions: $\phi = $ (a,d,g) 0.523, (b,e,h) 0.591 and (c,f,i) 0.598. Colour coding corresponds to spindle occupancy (a-c) pentagonal bipyramid occupancy (d-f). Probability distribution functions from these statepoints are also shown (g-i) for both spindle (green) and pentagonal bipyramid (blue) occupancy. Insets show the joint distributions for $O_{\mathrm{S}}$ and $O_{\mathrm{PBP}}$, and the corresponding Pearson correlation coefficients \textit{R} are shown. i) also indicates a configuration able to produce PBP occupancy in excess of 12, where neighbouring particles form a pentagonal ring (green bonds) with the central particle from the icosahedron, and the spindle (\jh{pink} particles) is formed by two particles from the icosahedral shell.
}
\label{figRender}
\end{figure}

Figure \ref{figRender} displays %to 
the spatial distribution of pentagonal bipyramids %PBP 
and spindle order for a range of state points. We see that for all state points, the spatial distributions of $O_{\mathrm{PBP}}$ and $O_{\mathrm{S}}$ are very similar. This is unsurprising: as discussed above, particles that contribute to many spindles, by definition, also contribute to many PBP. However, there are differences, and we see a more smoothly varying order parameter for $O_{\mathrm{PBP}}$ than for $O_{\mathrm{S}}$.%, while icosahedral particles are more clearly shown by $O_{\mathrm{Spindle}}$, as by definition the particles at a centre of an icosahedron maximise $O_{\mathrm{Spindle}}$.

We see that for a relatively low volume fraction ($\phi=0.523$, figure  \ref{figRender}(a,d)) the majority of particles are not detected in PBP, and those that are are spatially distinct. However, at higher volume fraction ($\phi=0.591$, figure \ref{figRender}(b,e)) a network of varying PBP and spindle membership can be observed. At higher volume fraction still ($\phi=0.598$, figure \ref{figRender}(c,f)) a similar network is found, but with richer PBP and spindle membership. Nevertheless, while some particles are detected in as many as 12 spindles (i.e., the centre of an icosahedron) there are still many particles that are not detected in any PBP or spindles. This is symptomatic of geometric frustration: while these five-fold symmetric structures can percolate through the sample, they are unable to tessellate, leaving regions poor in five-fold symmetry. We also note that the varying PBP field is reminiscent of studies of crystallisation that utilise bond-orientational order parameters such as $Q_{6}$ and $\psi_6$ \cite{tanaka2010}, as opposed to simple locally favoured structure occupancy, which is a binary (``on-off'') measure. Motivated by this observation, we proceed to investigate the change in PBP population with increasing volume fraction. %density.

We see distinct differences between different state points by considering the probability distribution function for pentagonal bypyramid occupancy (figure \ref{figRender}(g-i)). We see that for all densities the majority of particles are not detected in any spindles. However at high densities (figure \ref{figRender}(h,i)) the population in several spindles increases. We also note a second peak in the distribution for 12 spindles: these particles corresponds to the central particles in icosahedra. For pentagonal bipyramid occupancy we see similar behaviour with increasing volume fraction, albeit without a second peak in the distribution, and we also note PBP occupancy in excess of 12 at high densities. One might ask how a particle can contribute to more than 12 pentagonal bipyramids, exceeding its number of nearest neighbours. \jh{This is illustrated in the render 
in fig \ref{figRender}i, where an icosahedron formed by the white, purple, pink and dark green particles is joined by two neighbouring particles (light green). The central particle (purple) in the original icosahedron, in addition to two of the original shell particles (dark green) and two neighbouring particles (light green), can form a pentagonal ring (shown in green) for a spindle formed between two of the shell particles (pink) from the original icosahedron. This allows the original central particle to contribute to the 12 PBPs that form the original shell, and also contribute to the PBP with the two extra neighbours, allowing its PBP occupancy to exceed 12}. For a regular icosahedron the shell bond distance is 5\% longer than for the centre-shell bond distance \cite{royall2015physrep}, so the spindle and ring bond distances are reversed for an icosahedron-forming pentagonal bipyramid and a PBP that also includes \jh{extra} neighbouring particles. We also note that there is a strong positive correlation between $O_{\mathrm{S}}$ and $O_{\mathrm{PBP}}$ (see insets, figure \ref{figRender}(g-i)) at all state points, indicating that both order parameters are capturing similar structural features.

\begin{figure}
\centering
\includegraphics[width=0.99\textwidth]{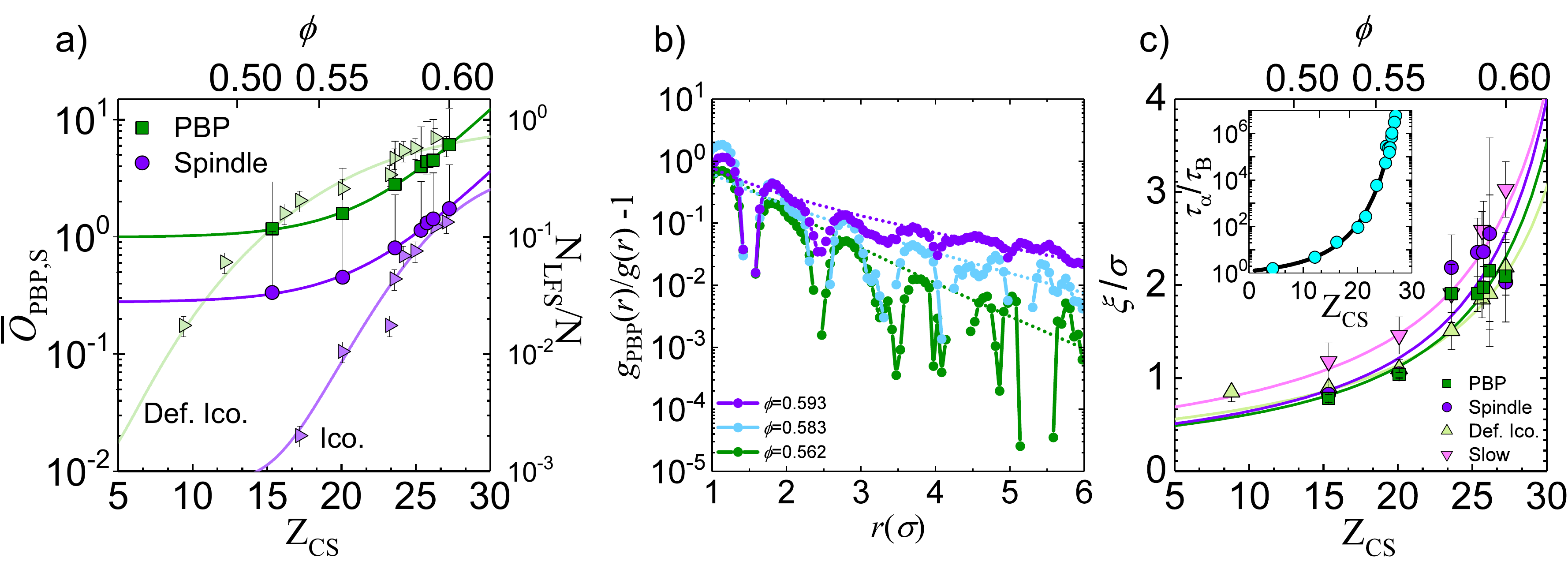}
\caption{\textbf{Pentagonal bipyramid populations and lengthscales}
(a) Occupancy of five-fold symmetric structures and populations of locally favoured structures as a function of pressure. Unlike the defective icosahedra, which saturates at high volume fraction, and the regular icosahedra, which has a negligible population until around $\phi_{\mathrm{MCT}}$ (reproduced from ref. \cite{hallett2018}), both spindle occupancy and pentagonal bipyramid occupancy increase monotonically at all state points. Lines are guides to the eye. For $O_{\mathrm{S}}$ and $O_{\mathrm{PBP}}$ only positive error bars are shown for clarity. (b) Example lengthscale fits for pentagonal bipyramid membership for state points $\phi = $ 0.562, 0.583 and 0.593. (c) Lengthscales associated with five-fold symmetric structures as a function of pressure. Lengthscales associated with both spindle occupancy and pentagonal bipyramid occupancy increases monotonically at all state points. Lines are fits to eq. \ref{eqChiara}, with parameters $\xi_0=58.2$ and $\theta=2.27\pm 0.11$ for $\xi_{\mathrm{S}}$ and $\xi_0=45.1$ and $\theta=2.23\pm 0.09$ for  $\xi_{\mathrm{PBP}}$. The lengthscales reported in reference \cite{hallett2018} are also shown, and produce similar scaling. Inset shows the Angell plot of $\tau_{\alpha}$ as a function of pressure. Datapoints are partly reproduced from \cite{hallett2018}. Solid line is a fit to VFT scaling (eq. \ref{eqVFTZ}), where $\phi_0=0.616$ $\pm$ 0.002 ($Z_0=31.1$).
}
\label{figLengthscales}
\end{figure}

We now consider the mean occupancy, and find that both spindle and PBP occupancies continue to increase with volume fraction (figure \ref{figLengthscales}a). This is notable because for a binary choice of locally favoured structure occupancy \cite{hallett2018}, defective icosahedra (the LFS for hard spheres) appears to saturate at high densities, while the population of regular icosahedra is negligible \textit{until} this saturation region. This measure of pentagonal bipyramid or spindle number appears to smoothly translate between the two structures. It is also possible to determine a lengthscale associated with these growing regions rich in pentagonal bipyramids or spindles. We determine the spindle or bipyramid weighted pair distribution function, as follows:

\begin{equation}
g_{a}(r)=\frac{\sum_{i,j} (w_{i,a} w_{j,a})\delta \left(r - \vert\overrightarrow{r_{ij}}\vert   \right)}{\left\langle w_{i,a}\right\rangle^{2}\pi r^{2}\Delta r \rho \left(N -1\right)}
\label{eqSlowgr}
\end{equation}
for \textit{a}= spindle or PBP, $\rho$ is the average number density,  $w_{i,a}$ is the occupancy and  $\left\langle w_{i,a}\right\rangle$ is the average occupancy. We then  extract a lengthscale by normalising against the particle-particle pair distribution function, and fit the decay using an Ornstein-Zernike envelope \cite{lacevic2003,dunleavy2015,tanaka2010} as follows:
\begin{equation}
\frac{g_{a}(r)}{g(r)}\sim \dfrac{1}{r} \mathrm{exp}\left(\dfrac{-r}{\xi_{a}}\right)
\label{eqLFSlength}
\end{equation}
where $\xi_{a}$ is the structural lengthscale for spindles or pentagonal bipyramids. Figure \ref{figLengthscales}b shows examples of this lengthscale fit. For increasing volume fraction the decay is slower, yielding a longer characteristic lengthscale. Figure \ref{figLengthscales}c shows the fitted lengthscales as a function of volume fraction and pressure. We choose to fit the characteristic increase in lengthscales using the following expression, inspired by random first order transition theory \cite{cammarota2009,royall2017}:

\begin{equation}
\xi[Z(\phi)]=\xi_0 \left( \frac{1}{Z_0-Z_\mathrm{CS}(\phi)} \right)^\frac{1}{3-\theta}.
\label{eqChiara}
\end{equation}

\noindent where $Z_0$ corresponds to the reduced pressure at which dynamical divergence is predicted by the VFT fit (see figure \ref{figLengthscales}c inset). For both spindles and pentagonal bipyramids, measured lengthscales markedly increase by almost a factor of three, with similar scaling (within error) to the previously reported lengthscales associated with slow dynamics and icosahedral order \cite{hallett2018}. By capturing the same structural features as defective icosahedra membership, but also capturing the details of the local five-fold (yet not fully icosahedral) environment, the lengthscales are generally slightly larger for both spindle and PBP occupancy than for defective icosahedra. This demonstrates the compatibility between the conventional ``locally favoured structures'' approach and the method adopted here. In addition to structure, global dynamics are indicated in the Angell plot (figure \ref{figLengthscales}c (inset)). We see that as populations and lengthscales associated with pentagonal bipyramids increase, the system slows down by almost 7 decades. In the following sections we will explore the interplay between slow dynamics and pentagonal bipyramid structure.

\subsection{Local dynamics}

\begin{figure}
\centering
\includegraphics[width=0.99\textwidth]{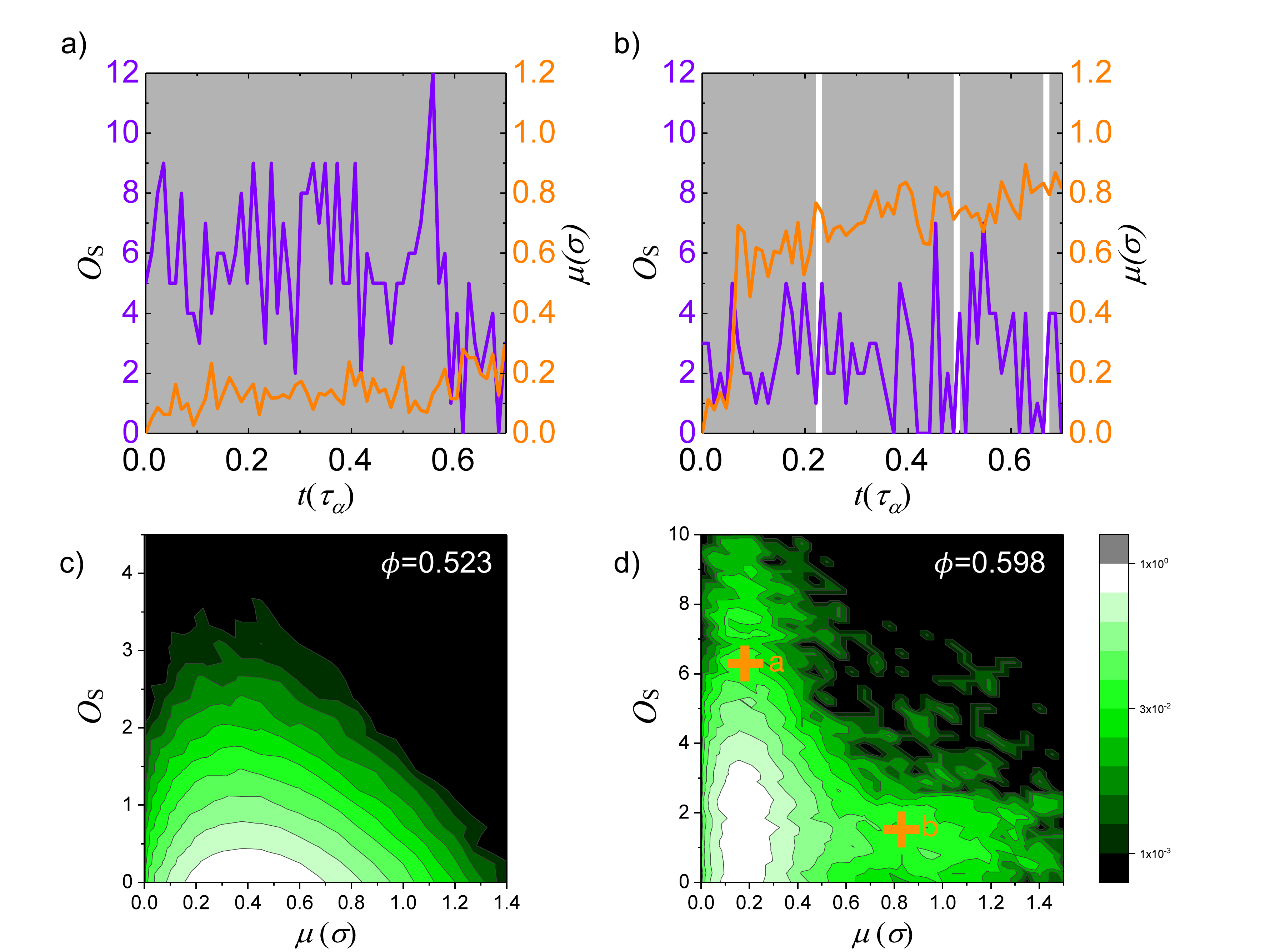}
\caption{\textbf{Pentagonal structure and slow dynamics}
Example displacements (orange) and spindle occupancy (purple) fluctuations for a spindle rich (a) and spindle poor (b) trajectory. Grey background indicates when the particle was also detected in a defective icosahedron, \jh{while white indicates non-detection in a defective icosahedron}. Joint probability distributions of mobility and persistency of spindle occupancy for state points c)0.523 and d) 0.598 over time interval $t \sim 0.5 \tau_{\alpha}$. Location of example trajectories shown in (a) and (b) are indicated in (d) by the orange symbols.
}
\label{figHeatmap}
\end{figure}

\begin{figure}
\centering
\includegraphics[width=0.7\textwidth]{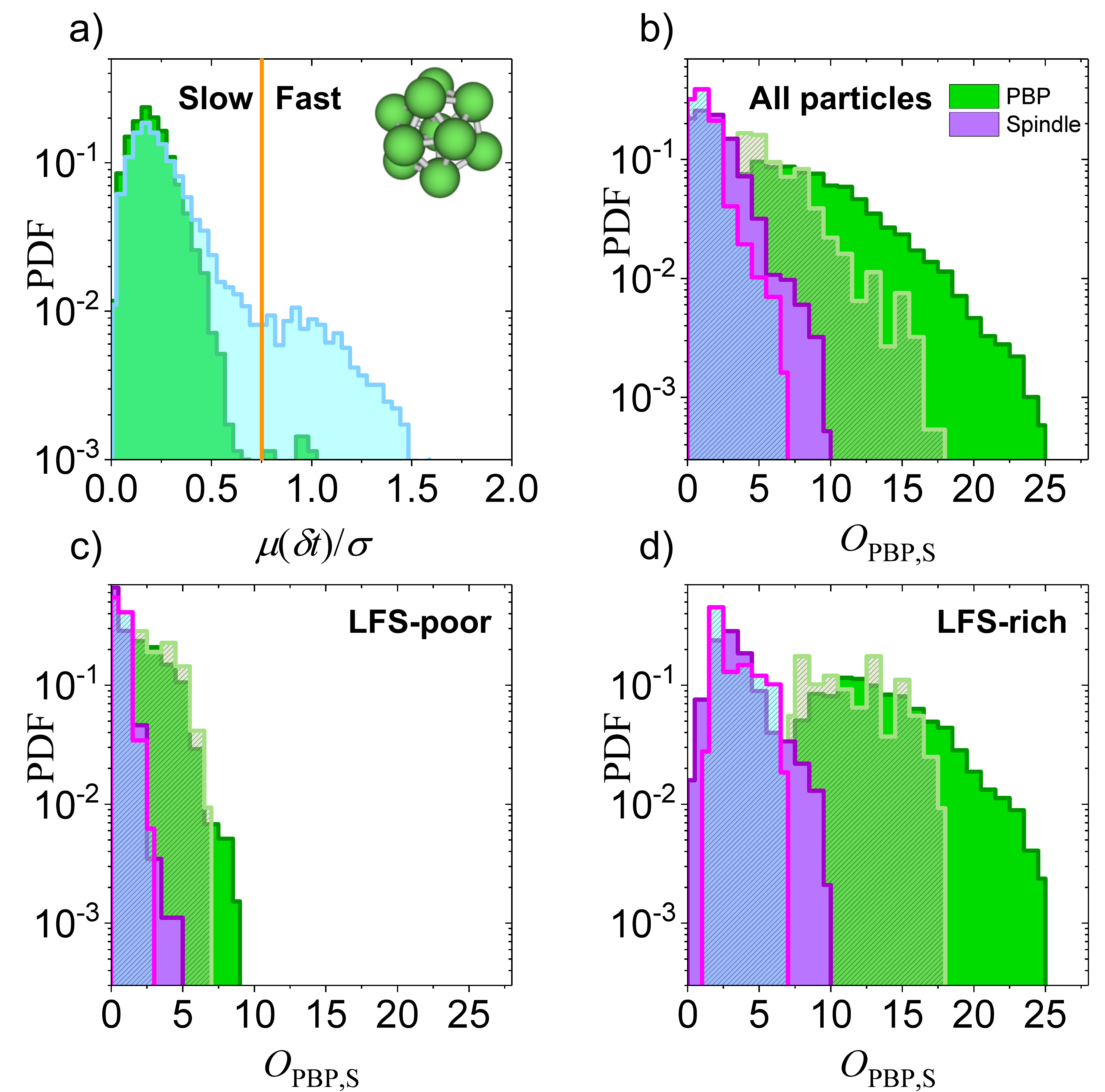}
\caption{\textbf{Defective icosahedra and mobility}
a) Distribution of mobility for trajectories rich in defective icosahedra locally favoured structures (green) and structurally poor trajectories (blue) for $\phi = 0.598$ for time interval $t \sim 0.5 \tau_{\alpha}$. Orange line indicates threshold for ``fast'' and ``slow'' particles. b) Distribution of $O_{\mathrm{S}}$ (purple) and $O_{\mathrm{PBP}}$ (green) for slow (solid colour) and fast (hashed colour) subsets. c) Distribution of $O_{\mathrm{S}}$ and $O_{\mathrm{PBP}}$ for slow and fast subsets for LFS-poor trajectories. d) Distribution of $O_{\mathrm{S}}$ and $O_{\mathrm{PBP}}$ for slow and fast subsets for LFS-rich trajectories.}
\label{fig10B}
\end{figure}

In order to determine the relationship between slow dynamics and local order, we now investigate the mobility of trajectories that persistently occupy many pentagonal bipyramids and spindles. We define the mobility \jh{$\mu$} as the displacement over a time interval $\sim 0.5 \tau_{\alpha}$ \cite{hallett2018}. Figures \ref{figHeatmap}c and \ref{figHeatmap}d show interpolated histograms of particle mobility and spindle occupancy for volume fractions 0.523 and 0.598 respectively. While the more weakly supercooled sample shows little correlation between spindle occupancy and mobility, the higher volume fraction sample shows a highly mobile, structurally poor subset and an immobile, structurally rich subset. To investigate these subsets in more detail, we show the displacements and corresponding spindle occupancy for example trajectories in figure \ref{figHeatmap}a and b, corresponding to positions marked in figure \ref{figHeatmap}d. We also indicate timesteps where the trajectory was detected in a defective icosahedron by shading the background, and note that, despite the differing spindle numbers and mobility between the example trajectories, they are both persistent in defective icosahedra -- a metric previously shown to favour slow dynamics. Inspired by this observation, we now pose the question ``Are pentagonal bipyramids a better predictor of slow dynamics than defective icosahedra?''. To phrase this slightly differently, can pentagonal bipyramid or spindle membership be used to identify the small subset of particles which are persistent in defective icosahedra, yet are also fast moving? To answer this, we determine the mobility of structurally rich particle trajectories that persist in defective icosahedra, and those that do not, following \cite{hallett2018} (where persistency is the fraction of steps during a trajectory where a particle is detected in a particular LFS). As described previously, we see that ``LFS-rich'' particles are generally slow moving, while ``LFS-poor'' particles can be either slow or fast moving, with a distinct second peak in the distribution (figure \ref{fig10B}a). We then sub-sample for ``fast'' and ``slow'' particles by using the minima between the slow and fast distributions for the LFS-poor trajectories ($\mu \sim 0.75 \sigma$ - indicated in figure \ref{fig10B}a) as a cutoff and investigate the pentagonal bipyramidal occupancy for these subsets. Figure \ref{fig10B}b shows the distribution of $O_{\mathrm{S}}$ (purple) and $O_{\mathrm{PBP}}$ (green) for slow (solid colour) and fast (hashed colour) subsets. We see lower $O_{\mathrm{S}}$ and $O_{\mathrm{PBP}}$ for fast moving particles than for slow moving particles. This is not surprising: we have already shown that fast moving particles are generally structurally poorer than slow moving particles (figure \ref{figHeatmap}). However, we now perform the same analysis for LFS-poor and LFS-rich particles and observe marked differences. For LFS-poor trajectories (figure \ref{fig10B}c) we see little distinction between fast and slow trajectories: in both instances the $O_{\mathrm{S}}$ and $O_{\mathrm{PBP}}$ are significantly lower than for the total population \jh{and the mean values (see table \ref{tableLFS}) are essentially the same}. However, for LFS-rich trajectories (figure \ref{fig10B}d) we see that fast moving particles generally have lower (approximately 15$\%$) $O_{\mathrm{S}}$ and $O_{\mathrm{PBP}}$ than slow moving particles \jh{(see table \ref{tableLFS})}. Physically, what does this mean? In a sense the lower occupancy is capturing either particles that are in isolated LFS, or those at the edge of icosahedral domains - which are not \jh{weighted differently when considering} simple LFS membership \jh{but are more able to move than, for example, particles in the core of an icosahedral domain}. It should be noted that for the sample described here, the vast majority ($> 99\%$) of LFS-rich particles are also slow moving, \jh{without needing to invoke any extra order parameter to identify them}. 
\jh{However, pentagonal bipyramid occupancy can be a useful tool to identify exceptional particle arrangements that are LFS-rich yet mobile}, or to identify the persistence and propagation of icosahedrally rich domains in a more informative way than simply through LFS-identification.
%However, pentagonal bipyramid occupancy can be a useful tool to identify characteristic structural signatures of exceptional particle arrangements \jh{that evade the simpler LFS order parameter}

%enumerate this? %mean and SD for structured/unstructured. Show figure?

\begin{table}
\caption{\label{tableLFS}Mean and standard deviation (in brackets) of LFS (defective icosahedra) occupancy and spindle and pentagonal bipyramid occupancy distributions, for fast and slow subsets obtained for all, LFS-rich and LFS-poor particles as determined from figure \ref{fig10B}.}
\begin{indented}
\lineup
\item[]\begin{tabular}{@{}llll}
\br
Subset & &  Slow &  Fast \\

\mr
&$\bar{O}_{\mathrm{LFS}}$&\0$0.80 (0.20) $&\0$0.66 ( 0.21) $\\
All&$\bar{O}_{\mathrm{S}}$&\0$2.3 (1.6) $&\0$1.6 (1.1) $\\
&$\bar{O}_{\mathrm{PBP}}$&\0$8.2 (4.6) $&\0$5.6 (2.9) $\\
\mr
&$\bar{O}_{\mathrm{LFS}}$&\0$0.49 (0.16) $&\0$0.47 (0.17) $\\
LFS-Poor&$\bar{O}_{\mathrm{S}}$&\0$1.04 (0.77) $&\0$1.06 (0.53) $\\
&$\bar{O}_{\mathrm{PBP}}$&\0$4.1 (1.6) $&\0$4.0 (1.5) $\\
\mr
&$\bar{O}_{\mathrm{LFS}}$&\0$0.99 (0.01) $&\0$0.98 (0.01) $\\
LFS-Rich&$\bar{O}_{\mathrm{S}}$&\0$ 4.0 (1.6) $&\0$3.5 (1.3) $\\
&$\bar{O}_{\mathrm{PBP}}$&$13.6 (3.5) $&$11.8 (2.9) $\\

\br
\end{tabular}
\end{indented}
\end{table}

\subsection{Dynamics and structural transitions}

\begin{figure}
\centering
\includegraphics[width=0.99\textwidth]{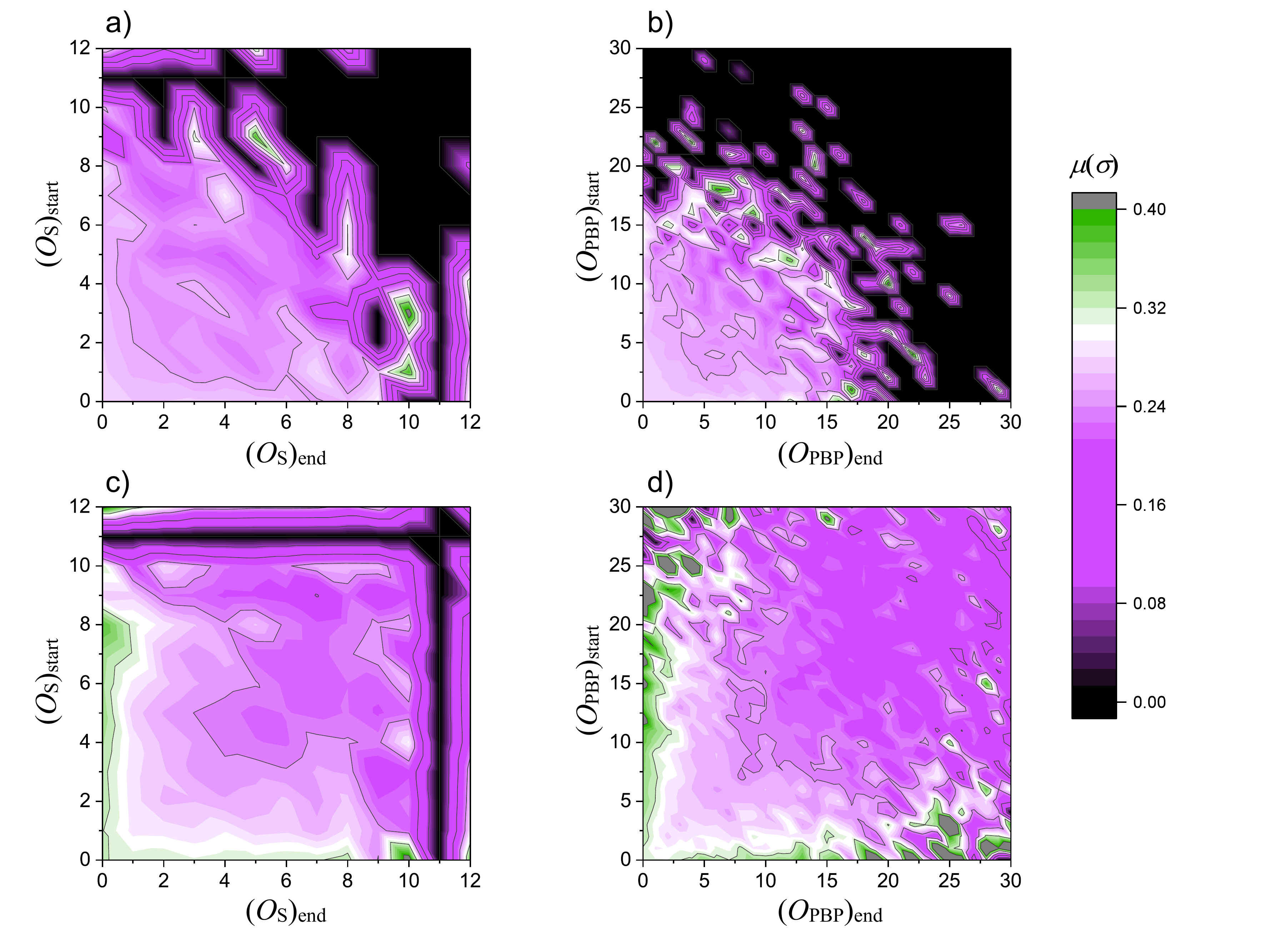}
\caption{\textbf{Mobility and five-fold structure transitions}
Map of mobility associated with specific structural transitions between spindle and bipyramid occupancies, over experimental timestep $t$
at state points $\phi = $ 0.562 (a,b, $t \sim 0.05 \tau_{\alpha}$) and $\phi=$ 0.598 (c,d $t \sim 0.05 \tau_{\alpha}$).}
\label{figTransitions}
\end{figure}

Figure \ref{figTransitions} shows the relationship between mobility and changing PBP structure over a single imaging step at high volume fraction. By considering the change in spindle number (or number of occupied PBPs) over this small time interval (typically $<0.05 \tau_{\alpha}$) we can associate this change in structure with large or small displacements. At weak supercooling (figure \ref{figTransitions}a,b) we see that both spindle and PBP occupancy show little clear correlation between mobility and structural transition. However, at deeper supercooling (\ref{figTransitions}c,d) we observe that in general transitions that conserve spindle or PBP occupancy result in small displacements, as expected for a structure that has been associated with slow dynamics but also note that transitions from a low spindle number (i.e., less locally structured) to high spindle number, or \textit{vice versa}, correspond to the largest measured displacements (figure \ref{figTransitions}c).
This signal is also present for PBP occupancy (figure \ref{figTransitions}d), although it is less pronounced than for spindle occupancy. This could emerge naturally as a consequence of the caging effect of these icosahedrally ordered structures: that is to say, in order to sufficiently disrupt these structures, large displacements of multiple particles are necessary. Also note that in figure \ref{figTransitions}a,c we observe that no particles were detected in 11 spindles, which we also see in the occupancy distribution in figure \ref{figRender}g-i. This indicates that, should a particle leave the surface of an icosahedron (where the central particle occupies 12 distinct spindles), either it is immediately replaced by a different particle, maintaining 12 spindles with the central particle, or the shell of particles around the central particle relaxes to a different structure, reducing the number of pentagonal rings and further reducing the number of detected spindles to 10 or fewer. This supports the claim that persistence in pentagonal bipyramidal motifs correlates with slow dynamics, but also reveals that ``structure breaking'' and ``structure making'' events can result in notable particle rearrangements.

%figure of lengthscales - roll in with populations as well?

%\subsubsection{Structural similarity}
%\jh{
%The results outlined above demonstrate that five-fold symmetric pentagonal bipyramid structures become increasingly important during vitrification. One appealing aspect of this characterisation is that, unlike for icosahedra or defective icosahedra, for which the discrimination rules are strict, pentagonal bipyramids can be arranged in many different ways yet produce the same distribution of spindle and PBP occupancy. Nevertheless, there should be similarities in the way particles pack in regions of high PBP occupancy which, are of greater extent than and not necessarily described by perfect icosahedra, are nevertheless characteristic of a deeply supercooled liquid.

%To quantify this, we utilise the static overlap: DEFINE HERE.

%We thus calculate the static overlap for randomly sampled regions containing \textit{N} particles, and then determine the distribution of these overlaps for different N.

%}

\section{Conclusions}
In this paper we explore the suitability of a novel five-fold symmetric order parameter to investigate the role of local structure in dynamic arrest. %In this paper we use a novel method to explore 
By using occupancy of pentagonal bipyramidal structures as a measure of local ``icosahedral quality'', we have unveiled new characteristic structural signatures of slow dynamics, while also reproducing many of the observations shown through simple membership of locally favoured structures. We demonstrate that, while spindle and pentagonal bipyramids number are finer variables than simple %LFS 
\jh{occupancy of locally favoured structures,} dynamic correlations and measured lengthscales are highly compatible between the two approaches. Intriguingly we show that both PBP and spindle number continue to increase over the range of densities explored here and capture the growing five-fold symmetry of the vitrifying liquid. We also show that large instantaneous displacements are characteristic of sudden transitions in spindle and pentagonal bipyramid occupancy number. Finally, we make a direct comparison between the pentagonal bipyramid approach and locally favoured structures and distinguish exceptional cases of fast-moving, yet structurally-rich particles from slow-moving structurally-rich particles through their pentagonal bipyramid signature.

This approach can be readily adopted in structural studies of supercooled liquids, either from experiments or in simulations as an additional order parameter to characterise the system. While this approach can yield more quantitative information than the ``standard'' locally favoured structures approach, we are gratified to see a high degree of compatibility between the two approaches. Indeed, one can think of traditional icosahedra and defective icosahedra identification as equivalent to a measure of pentagonal bipyramidal or spindle occupancy with a neighbour-specific threshold. On the other hand, the approach adopted here exclusively prioritises five-fold symmetry at the expense of other local structure. We therefore propose a combined approach for future particle resolved studies: To use the topological cluster classification (or alternative local structure identification) as a qualitative analysis of the system structure, supplemented by this approach to quantify icosahedral ordering.

%\jh{In this paper we explore the role of pentagonal bipyramids in dynamic arrest. We demonstrate that, while spindle and PBP number are more continuous variables than simple LFS occupancy, dynamic correlations and measured lengthscales are highly compatible between the two approaches. Intriguingly we show that both PBP and spindle number continue to increase over the range of densities explored here and capture the growing five-fold symmetry of the vitrifying liquid. We also show that large instantaneous displacements are characteristic of sudden transitions in spindle and pentagonal bipyramid occupancy number. Finally, we make a direct comparison between the pentagonal bipyramid approach and locally favoured structures and distinguish fast-moving, yet structurally-rich particles from slow-moving structurally-rich particles through their pentagonal bipyramid signature.}

\section{Acknowledgements}
JEH thanks Martin van Hecke, whose insightful question prompted this investigation. \paddyspeaks{CPR acknowledges Patrick Charbonneau for his general wisdom.}
CPR acknowledges the Royal Society, and the Kyoto University SPIRITS fund. JEH, FT and CPR acknowledge the European Research Council (ERC consolidator grant NANOPRS, project 617266) and the Engineering and Physical Sciences Research Council (EP/H022333/1) for financial support.

\newpage

%\bibliographystyle{unsrt}
%\bibliographystyle{iopart-num}
%\bibliography{highSparrow}

\end{document}